\documentclass{aa}  
\usepackage{natbib}
\bibpunct{(}{)}{;}{a}{}{,} 

\usepackage{soul}
\usepackage{color}
\usepackage{txfonts}
\usepackage{natbib}
\usepackage[squaren,Gray]{SIunits}

\newcommand{\gcc}{\mbox{ g cm$^{-3}$}}
\usepackage{bbold}
\usepackage{subcaption}
\usepackage{hyperref}

\usepackage{mwe}

\newcommand{\msun}{$M_\odot$\,}

\newcommand{\mic}{$\mu$m}

\newcommand{\apeak}{$a_{\rm{peak}}$\,}
\newcommand{\dvgrain}{$\delta \mathrm{v}_{a_{\mathrm{peak}}}$\,}
\newcommand{\St}{\mathrm{St}}

\newcommand{\simgt}%
        {\,\hbox{\lower0.6ex\hbox{$\sim$}\llap{\raise0.6ex\hbox{$>$}}}\,}
\def\simlt{\lower.5ex\hbox{\ltsima}}
\def\ltsima{$\; \buildrel < \over \sim \;$}
\def\simlt{\lower.5ex\hbox{\ltsima}}
\def\gtsima{$\; \buildrel > \over \sim \;$}
\def\simgt{\lower.5ex\hbox{\gtsima}}

\usepackage{graphicx}
\usepackage{txfonts}
\begin{document}

   \title{On the 3D time evolution of the dust size distribution in protostellar envelopes}

   \author{M.~Lombart\inst{\ref{inst1}}
            \and
            U.~Lebreuilly\inst{\ref{inst1}}
            \and
            A.~Maury\inst{\ref{inst1},\ref{inst2},\ref{inst3}}
    }

    \institute{
    Universit\'e Paris-Saclay, Universit\'e Paris Cit\'e, CEA, CNRS, AIM, F-91191 Gif-sur-Yvette, France\label{inst1}
    \\
    \email{maxime.lombart@cea.fr}
    \and
    Institute of Space Sciences (ICE), CSIC, Campus UAB, Carrer de Can Magrans s/n, E-08193 Barcelona, Spain\label{inst2}
    \and
    ICREA, Pg. Lluís Companys 23, Barcelona, Spain\label{inst3}
    }

   \date{Received XXX; accepted XXX}

 
  \abstract
   {Dust plays a fundamental role during protostellar collapse, disk and planet formation. Recent observations suggest that efficient dust growth may begin early, in the protostellar envelopes, potentially even before the formation of the disk. Three-dimensional models of protostellar evolution, addressing multi-size dust growth, gas and dust dynamics and magnetohydrodynamics, are required to characterize the dust evolution in the embedded stages of star formation.}
   {We aim to establish a new framework for dust evolution models, following in 3D the dust size distribution both in time and space, in MHD models describing the formation and evolution of star-disk systems, at low numerical cost.}
   {We present our work coupling the \texttt{COALA} dust evolution module using the Smoluchowski equations to \texttt{RAMSES} numerical computations, performing the first 3D MHD simulation of protostellar collapse including simultaneously polydisperse dust growth modeled by the Smoluchowski equation as well as dust dynamics in the terminal velocity approximation.}
   {We note that locally in the protostellar collapse the dust size distribution deviates significantly from the MRN due to dust growth. Ice-coated micron-sized grains can rapidly grow in the envelope and survive by not entering the fragmentation regime. The evolution of the dust size distribution is highly anisotropic due to the turbulent nature of the collapse and the development of favorable locations such as outflow cavity walls, which enhance locally the dust-to-gas ratio.}
   {We analyzed the first 3D non-ideal MHD simulations that self-consistently account for the dust dynamics and growth during the protostellar stage.  Very early in the lifetime of a young embedded protostar, micron-sized grains can grow, and locally the dust size distribution deviates from the MRN initial shape. This new numerical method opens the perspective to treat simultaneously gas/dust dynamics and dust growth in 3D simulations at a low numerical cost for several astrophysical environments. 
   }


   \maketitle
%

\section{Introduction}

The formation and early evolution of dust grains in young stellar systems represent a critical yet poorly understood component of star and planet formation processes. While dust grains constitute only $\sim$1\% of the interstellar medium's mass, they fundamentally influence protostellar evolution through their roles in radiative cooling, chemistry catalysis, and magnetic field coupling \citep{Draine2003, Zhao2016,Lebreuilly2023,Valluci2024}. 

Observations of the dust thermal emission at different frequencies allow to measure the dust emissivity $\beta$, a quantity that is key to computing dust masses from dust emission but is also widely used to measure the mean dust grain size in astrophysical environments and discuss, for example, dust evolution from submicronic solid particles towards pebbles and the formation of planetary systems \citep{Testi2014, Birnstiel2024}. Some pilot observational studies had suggested that young ($<1$ Myrs old) disks may show evolution of the dust emissivity consistent with an increase of the mean dust grain size \citep{Kwon2009, Miotello2014}. More recently, observations have revealed compelling evidence that dust evolution is more complex than expected and may begin earlier than previously assumed (maybe even starting during the prestellar phase), challenging traditional models that assumed significant grain growth only occurred in disks. For example, \citet{Dartois2024} modeled the CO$_2$ ice band observed by JWST in two line-of-sights in a dense cloud (N(H) $\sim 1-2 \times 10^{23} \, \rm{cm}^{-2}$, A$_v > 60$). They interpret the extended blue wing of this band as an excess of scattered light, and conclude that a fair fraction of the dust mass must be in grains $\sim 1$ micrometer in size to explain this spectral feature. At longer wavelengths, observational studies of the Taurus B211/B213 star-forming filament have shown a systematic decrease in dust emissivity index ($\beta$) from $\beta \approx 2$ in prestellar cores at scales $\sim 10\,000$ au, to $\beta < 1$ towards protostellar envelopes \citep{Ysard2013, Bracco2017}. Low dust emissivities $\beta$ have also been observed in the inner envelopes of embedded protostars, with high-angular resolution studies of the millimeter dust emission \citep{Galametz2019, Cacciapuoti2023}. All together, these studies reveal a variety of dust optical properties, while the lowest values of the dust emissivity $\beta$ can only be reproduced so far by including large dust grains, with $>100 \,\mu$m sizes, suggesting efficient grain aggregation could begin long before dust grains have entered the disk environment, in the dense cores and envelopes. 

Improving our modeling of the dust size distribution during the star-formation sequence requires us to address the magnetohydrodynamics (MHD) of gas and dust mixtures simultaneously with a full treatment of dust evolution. State-of-the-art dust evolution models employ distinct numerical strategies to balance computational feasibility with physical fidelity. Three-dimensional simulations are required to capture asymmetries in the evolution of the collapse. However, they are challenging and therefore often adopt a monodisperse or bi-disperse \footnote{Single or two grain sizes, potentially varying locally.} approximation to resolve dust growth \citep{Vorobyov2019,Elbakyan2020,Tsukamoto2021,Vorobyov2024,Vorobyov2025} or do not include the dust-gas differential dynamics \cite{Bate2022,Marchand2023}. We note the very recent effort of \citet{Bate2025} to couple dust dynamics with a full treatment of dust evolution with the Smoluchowski equation. They employed these methods for collapse calculation that resemble those presented in this article but with SPH methods. Conversely, several codes have been developed to solve the Smoluchowski equation across a wide span of grains size with the gas and dust dynamics in 1D, such as \texttt{DustPy} \citep{Stammler2022}, \texttt{shark} \citep{Lebreuilly2023,Valluci2024}, or in 2D, such as \texttt{LA-COMPASS} \citep{Drazkowska2019,Li2020,Ho2024}, \texttt{CUDISC} \citep{Robinson2024}, with a significant numerical cost. Recent efforts like the \texttt{TriPoD} framework \cite{Pfeil2024} aim to reduce this numerical cost by embedding parameterized size distributions in 2D hydrodynamic grids, though challenges persist in reconciling growth timescales with dynamical environments. However, incorporating accurate numerical solutions to the Smoluchowski equation in 3D multiphysics hydrodynamics simulations, is out of reach with the mentioned algorithms. This numerical challenge requires a new algorithm to efficiently solve the Smoluchowski equation. Thanks to the newly developed \texttt{COALA} code it is now possible to fully solve the Smoluchowski equation with a wide span of grain size on-the-fly in 3D simulations that also follow dust dynamics. In the present study, using a coupling between \texttt{COALA} and the adaptive mesh-refinement code \texttt{RAMSES}, we present new simulations beyond the current state-of-the-art that follow non-ideal MHD protostellar collapses with a full solving of the Smoluchowski equation joint with a treatment of dust dynamics in the terminal velocity approximation. 
   

\section{Methods}

In this work, we perform numerical 3D MHD simulations of protostellar collapses using the code \texttt{RAMSES} \citep{Teyssier2002}. In particular, we employ its extension to non-ideal MHD \citep{Masson2012}. We follow the evolution of a gravitationally unstable prestellar core of $2.5~M_{\odot}$ with an initial radius $\sim 8000 $ au \footnote{this corresponds to a thermal-to-gravitational energy ratio of $\alpha=0.35$} and an initial turbulent Mach 0, or 1 (denoted $\mathcal{M}_0$ and $\mathcal{M}_1$). Initial conditions are similar to models shown in \citet{Maury2019} and \citet{Valdivia2022}. The temperature of the cloud is computed assuming a barotropic equation of state as per \cite{Lebreuilly2020}, this assumes isothermality below $10^{-13} \gcc$ and adiabaticity with an adiabatic index of $5/3$ above. We use the adaptive-mesh refinement capability of \texttt{RAMSES} to follow the gravitational collapse. The coarse grid has a resolution of 1100 au and is refined imposing at least 10 points per Jeans length to avoid artificial fragmentation \citep{Truelove1997} and reaches a minimum cell size of 4.3 au in the densest regions. 

\subsection{Dust growth with COALA}
\label{sec:coala}
We incorporate a full treatment of the dust coagulation by solving on-the-fly the Smoluchowski equation for spherical grains of single composition in the conservative form  \citep{Tanaka1996}  using the \texttt{COALA} code \citep{Lombart2021}. Assuming a mass range $[m_{\mathrm{min}},m_{\mathrm{max}}]$ for the dust grains, the equation writes as
\begin{equation}
\begin{aligned}
&\frac{\partial g(m,t)}{\partial t} + \frac{\partial F_{\mathrm{growth}}[g](m,t)}{\partial m} = 0,\\
&F_{\mathrm{growth}}[g](m,t) \equiv \\
& \qquad \int_{m_{\mathrm{min}}}^{m} \int_{m-m_1+m_{\mathrm{min}}}^{m_{\mathrm{max}}-m_1 +m_{\mathrm{min}}} \frac{K(m_1,m_2)}{m_2} g(m_1,t)g(m_2,t) \mathrm{d}m_2 \mathrm{d}m_1,
\end{aligned}
\label{eq:smolu_cons}
\end{equation}
where $m \in [m_{\mathrm{min}},m_{\mathrm{max}}]$ is the mass of dust grains, $g(m,t)$ is the dust mass distribution, and $K(m_1,m_2)$ the collision kernel between the two grains of mass $m_1$ and $m_2$. We use the ballistic kernel $K(m_1,m_2)=\sigma(m_1,m_2) \Delta v(m_1,m_2)$ where $\sigma$ is the collisional cross-section and $\Delta v$ is the differential velocity between grains. In this study, the sources of grain-grain differential velocities are brownian motion, hydrodynamical drift and gas turbulence (see details in Appendix~\ref{sec:dust_dv}). We use the high accuracy of the \texttt{COALA} code to follow the evolution of the dust size distribution with a low number of dust size bins. The time solver implemented in \texttt{COALA} is a Strong Stability Preserving Runge-Kutta (SSPRK) third order \citep{Gottlieb2009}. In each spatial cell, we use $\mathcal{N}>1$ dust species for which the set of dust mass density $\left\{\rho_k\right\}_{1 \leq k \leq \mathcal{N}}$ defines the dust size distribution. The dust mass distribution $g$ in Eq.~\ref{eq:smolu_cons}
is linked to $\rho_k$ such as
\begin{equation}
\rho_k = \int_{I_k} g(m,t) \mathrm{d}m,
\end{equation}
with $I_i$ the size of bin $k$. In this study, the code \texttt{COALA} gives the evolution of the mean value of the function $g$, noted $\overline{g}$, meaning that the piecewise constant approximation is used. Therefore, the relation between $\rho_k$ and $\overline{g}_k$ is directly obtained $\overline{g}_k = \rho_k/\Delta m_k $. The accuracy on the approximation of $g$ depends on the number of dust bins $\mathcal{N}$. In Appendix~\ref{sec:benchmark_coala}, \texttt{COALA} is benchmarked with exact solutions in the setup given in Sect.~\ref{sec:dust_init}.

\subsection{Dust dynamics with RAMSES}
\label{sec:ramses}
The dust dynamics is treated by the terminal velocity approximation solver of \cite{Lebreuilly2019}. In short, this approach  considers dust species as separate phases in a monofluid composed of the full gas-dust mixture with a density $\rho$ and velocity $\vec{v}$. Each dust phase $k$ of density $\rho_k$ requires to solve a mass conservation equation with a velocity $\vec{v}_k = \vec{v} + \vec{w}_k$ where $\vec{w_k}$ is given by the formula 
\begin{eqnarray}
\label{eq:wkB}
 \vec{w_k} = \left[\frac{\rho}{\rho-\rho_k}t_{\mathrm{s},k}- \sum_{l=1}^{\mathcal{N}} \frac{\rho_l}{\rho-\rho_l} t_{\mathrm{s},l}\right] \frac{\nabla P_{\mathrm{th}} - (\nabla \times \vec{B})\times \vec{B}}{\rho},
 \end{eqnarray}
 where $P_{\mathrm{th}}$ is the gas thermal pressure, $\vec{B}$ is the magnetic field and $t_{\mathrm{s},k}$ is the grain stopping time. In the \cite{Epstein1924} regimes, for a grain of size $S_{\mathrm{grain},k}$ and intrinsic density $\rho_{\mathrm{grain},k}$, it writes as 

 \begin{eqnarray}
     \label{eq:tstopkw}
   t_{\mathrm{s},k} \equiv \frac{\rho_{\mathrm{grain},k}}{\rho}\frac{S_{\mathrm{grain},k}}{w_{\mathrm{th}}},
\end{eqnarray}
this regime is valid for small grains of the ISM provided that their drift velocity is small compared to the gas thermal speed $w_{\mathrm{th}}=\sqrt{\frac{8 k_{\mathrm{B}} T}{\pi \mu_{\mathrm{g}}m_{\mathrm{H}}}}$.

\subsection{Coupling between \texttt{RAMSES} and \texttt{COALA}}
\label{sec:coupling}
The dust growth equation is added as a source term in the continuity equation for each dust fluids \citep[see][]{Lebreuilly2023} as
\begin{equation}
\left\{
\begin{aligned}
&\frac{\partial \rho_k}{\partial t} + \mathbf{\nabla} \cdot [\rho_k \vec{v}_k]= \Lambda_{k,\mathrm{growth}},\\
& \Lambda_{k,\mathrm{growth}} \equiv \left. \frac{\partial \rho_k}{\partial t} \right|_{\mathrm{growth}} \approx \frac{ \mathrm{d} \overline{g}_k}{\mathrm{d} t} \Delta m_k,
\end{aligned}
\right.
\end{equation}
where $g$ is the unknown variable in the Smoluchowski equation (see Sect.~\ref{sec:coala}). Then, an ordinary operator splitting is applied to solve these continuity equations. First, the dynamic of the dust fluids is solved with the code \texttt{RAMSES}, and then the dust growth source term is solved by the code \texttt{COALA}. This is illustrated by using an Euler forward time scheme 
\begin{equation}
\begin{aligned}
&\rho_k^* = \rho_k(t_n) - \Delta t \; (\mathbf{\nabla} \cdot [\rho_k \vec{v}_k]) \rightarrow \texttt{RAMSES},\\
& \rho_k(t_{n+1}) = \rho_k^* +  \Delta t \; \Lambda_{k,\mathrm{growth}} \rightarrow \texttt{COALA},
\end{aligned}
\end{equation}
where $\Delta t$ is the hydrodynamic time-step. In this study, magnetic resistivities of grains are not evolving with changes in the dust size distribution.

\subsection{Dust initial conditions}
\label{sec:dust_init}
For this work, the binning is done logarithmically with $\mathcal{N}=40$ between $s_{\mathrm{min}} = ~5~\nano\meter$ and $s_{\mathrm{max}} = 1~\centi\meter$. We set-up the dust-to-gas ratio of each bin assuming a MRN-like \citep{Mathis1977} power law distribution between $s_{\mathrm{min}} = ~5~\nano\meter$ and $s_{\mathrm{cut}} \sim  270~\nano\meter$, which is the right interface of the bin $N_{\mathrm{cut}}=11$, including the grain size $250~\nano\meter$. For bins with $i>N_{\mathrm{cut}}$, we set an initial dust ratio with a very low value of $10^{-20}$. For each bin $k$ between 1 and $N_{\mathrm{cut}}$ we have an initial dust ratio 
\begin{equation}
    \epsilon_{k,0}= \epsilon_0 \left[ \frac{S_{k,+}^{4+m}-S_{k,-}^{4+m}}{S_{\mathrm{+,N_{\mathrm{cut}}}}^{4+m}-S_{\mathrm{-,1}}^{4+m}}\right],
\end{equation}
assuming a MRN with $m=-3.5$. $S_{k,-}$ and $S_{k,+}$ are the grain size at left and right interface of bin $k$. As explained in \cite{Lebreuilly2019}, the grain size that is used to dynamically evolve the bin $k$ is taken to be $\sqrt{S_{k,-} S_{k,+}}$. Note that, we consider in this study that all grains have the same intrinsic density $\rho_{\mathrm{grain}}= 2.3~\gram\, \centi\meter^{-3}$. 

\section{Results}
\label{sec:results}
\begin{figure*}
    \centering
    \includegraphics[width=\textwidth,trim={0cm 2cm 0cm 2cm},clip]{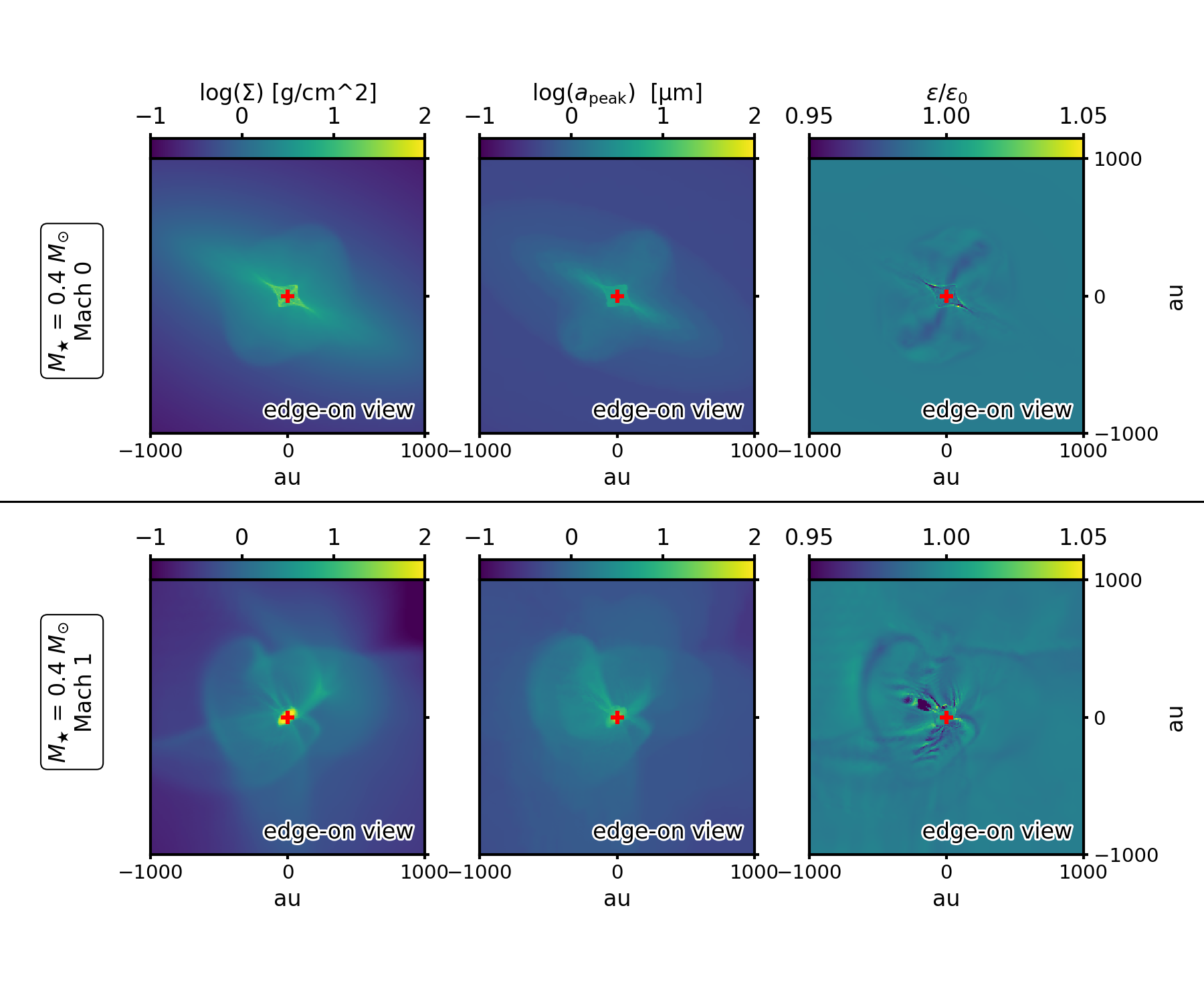}
    \caption{Snapshots taken from the \texttt{RAMSES} MHD models ($\mathcal{M}_0$ and $\mathcal{M}_1$) of a collapsing protostellar envelope (initial parameters can be found in the text) with dust grain size evolution computed by \texttt{COALA}. The maps show a 2000-au zoom-in of the model when the central star has accreted 0.4\msun \,from the initial 2.5\msun \,core. The red cross represents the position of the star. The left column shows the total column gas density $\Sigma$. The middle column shows the average grain size at the peak of the size distribution $a_{\mathrm{peak}}$, weighted by the gas density along the line of sight (see text for further details). Similarly, the right column shows the average dust enrichment (ratio between the evolved dust-to-gas ratio $\epsilon$ and the initial dust-to-gas ratio $\epsilon_0$) weighted by the gas density along the line of sight. On the top row, the square patterns near the sink are not numerical artifacts but result from the weighted average of the quantities along the line of sight.}
    \label{fig:maps}
\end{figure*}

The \texttt{RAMSES} model of a collapsing protostellar envelope with dust growth computed by \texttt{COALA} presented here is the first 3D MHD model of dust evolution during the disk formation stage that self-consistently accounts for both the gas and dust dynamics (multiple dust sizes) and the dust size evolution modeled by the Smoluchowski equation.

To exclude the impact of the grid boundaries, we carry out the analysis of the dust evolution in the inner half of the initial core extent, i.e. in the inner envelope at radii $<4000$ au (see Fig.~\ref{fig:full_size_map}). We ran two models with and without turbulence, referred as $\mathcal{M}_0$ and $\mathcal{M}_1$, and follow their evolution until the protostar accreted one half of the total core mass (0.85 \msun in the sink). From these two models, we selected a snapshot after the formation of the star when $M_{\star} = 0.4$ \msun, i.e. $t_{0.4\, M_{\odot}} = 118.1\,\mathrm{kyr}$ for model $\mathcal{M}_1$ and $t_{0.4 M_{\odot}} = 102.4\,\mathrm{kyr}$ for model $\mathcal{M}_0$. This evolutionary stage is representative of a young embedded protostar. In the following, the disk region is defined as the region where the gas number density is greater than $10^9\, \mathrm{cm^{-3}}$. For these snapshots, the disk radius is $\lesssim 80$ au: thus, radii $>100$ au probe exclusively the envelope material.

Figure \ref{fig:maps} shows the two models $\mathcal{M}_0$ and $\mathcal{M}_1$, taken when the central stellar object has accreted 0.4 \msun. 
The models are seen edge-on, e.g. with the disk rotation axis in the plane of the sky. The left column shows the map of total \{gas+dust\} column density $\Sigma$, computed as the density integrated along the line-of-sight. The middle column shows the map of the grain size $a_{\mathrm{peak}}$ at the peak of the size distribution (in each cell, the grain size $a_{\mathrm{peak}}$ is computed as the average size in the size-bin showing the largest mass fraction in the dust grain size distribution), averaged along the line-of-sight and weighted by the local gas density $\rho$, see Appendix \ref{sec:av_quant}. The right column shows the dust enrichment, defined as the ratio between the evolved dust ratio $\epsilon=\rho_{\mathrm{dust}}/(\rho_{\mathrm{gas}}+\rho_{\mathrm{dust}})$ and the initial dust ratio $\epsilon_0$, averaged along the line-of-sight and weighted by the gas density along the line of sight, see Appendix \ref{sec:av_quant}. The boxy structures around the star for the model $\mathcal{M}_0$ appear only when integrating the density along the line-of-sight at scales greater than 1000 au, and are not seen in individual model slices. Moreover, they appear only in the model without turbulence.These structures are not numerical artifacts from the simulation, but are due to a visualization effect. Therefore, these patterns are not a problem for the physical interpretation of the results.

Both models show significant dust evolution, already at the early times, in the disks but also in the protostellar envelopes. 
For example, for the model $\mathcal{M}_1$, accounting for both disk and envelope contributions in a sphere with radius 4000 au around the star, the grains with size between 0.188 and 0.270 $\mu$m initially account for $~\sim 20\%$ of the total dust mass, while at $M_{\star} = 0.4$ \msun after the collapse starts, these grains have been depleted to produce larger grains and represent only $~\sim 0.1\%$ of the dust mass in the disk and $~\sim 13\%$ of the dust mass in the envelope. This depletion of the small grains is associated to an enrichment in large, micron-sized grains. For example, for the dust bin containing the grains $5 \,\mu$m, the dust mass enrichment is $\sim 10\%$ in the disk and $\sim 0.6 \%$ in the envelope. 
We note that the denser the material is, the larger the grains. This is expected, as the material becomes denser, the collision rate between grains increases. As shown in Appendix.~\ref{sec:scaling}, assuming a simple spherical collapse we can show that the peak of the dust size distribution scales $\propto\sqrt{\rho}$. 
\begin{figure*}
    \centering    
    \includegraphics[width=\textwidth,trim={0cm 2cm 0cm 2cm},clip]{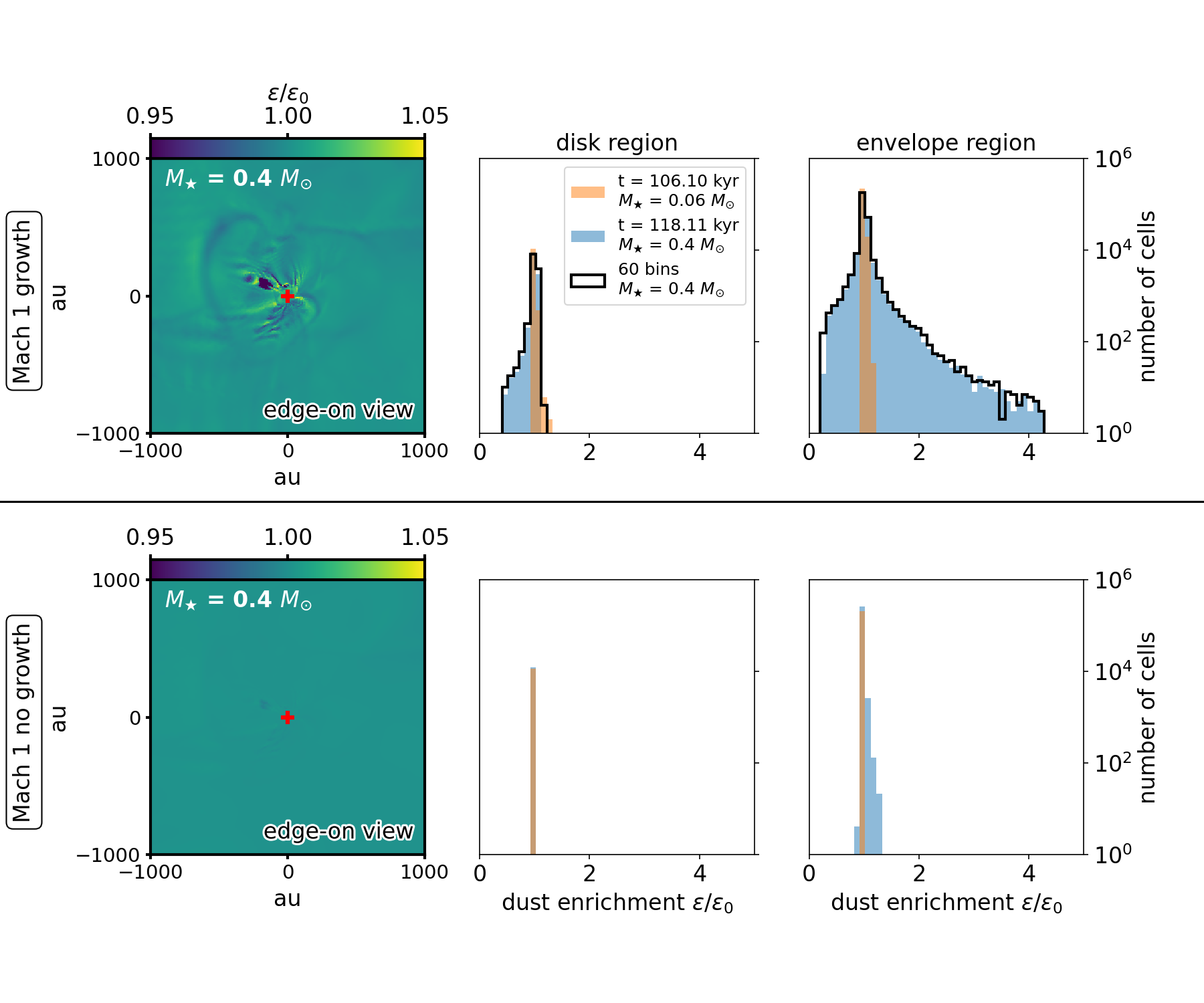}
    \caption{Snapshots and histograms of the dust enrichment at two given times of the model $\mathcal{M}_1$ with (top row) and without (bottom row) dust growth. The histograms are obtained with the dust enrichment calculated in each cell in a box of side 2000 au around the star. The two times are chosen when $M_{\bigstar}=0.06\, M_{\odot}$ (orange) and $M_{\bigstar}=0.4\, M_{\odot}$. Dust growth and dynamics enhance larger local variations of the dust enrichment, greater than 2 in the envelope compared to the dust dynamics alone. The black line shows the histograms for the $\mathcal{M}_1$ model with dust growth and 60 dust bins, highlighting the numerical convergence reached with 40 bins simulations. The red cross represents the position of the star. The variance value $\sim 0.01$ of the histogram explains the low variation observed on the map for the model $\mathcal{M}_1$ when $M_{\bigstar}=0.4\, M_{\odot}$.}
    \label{fig:dust_enrichment_hist}
\end{figure*}

Our work in 3D shows that the dust evolution is highly anisotropic in both models $\mathcal{M}_0$ and $\mathcal{M}_1$, with preferred locations for producing large dust grains either associated to the development of higher density structures due to a turbulent infall ($\mathcal{M}_1$), or due to the development of outflow cavity walls ($\mathcal{M}_0$). Figure~\ref{fig:M1_g_ng}, in Appendix~\ref{sec:M1_g_ng}, shows the anisotropy of the dust evolution by the comparison of the simulation for the model $\mathcal{M}_1$ with and without dust growth. Moreover, our models show that implementing the size evolution of the dust grains during the collapse amplifies local variations of the dust ratio. Figure~\ref{fig:dust_enrichment_hist} compares the evolution of the dust enrichment for models $\mathcal{M}_1$ with and without growth. The histograms are obtained with the dust enrichment calculated in each cell in a box of side 2000 au around the star. With dust growth, the dust enrichment can reach a value greater than 2 in the envelope compared to the disk region. However, only in a few cells the dust enrichment has high value: the variance of the histogram has low value $\sim 0.01$ for the model $\mathcal{M}_1$ when $M_{\bigstar}=0.4\,  M_{\odot}$. This low variance explains the small variation observed in the map in Figs.~\ref{fig:maps},\ref{fig:dust_enrichment_hist}.

This does not come out as a surprise since the Stokes number, the ratio between the stopping time and the free-fall time, is proportional to the grain size. As shown in \cite{Lebreuilly2020}, larger Stokes number produce large dust-to-gas ratio variations. It is interesting to point out that the Stokes number is roughly constant because of the scaling of the grain size with the density. This explains, why variations of dust-to-gas ratio remain relatively small even when incorporating dust growth. Indeed these variations essentially depend on the initial choice of grain size and the Stokes number of the peak of the MRN distribution is $\ll 1$. We point out that the strongest dust-to-gas ratio variations are in the regions of strong electric current and are therefore not caused by pressure gradients. This effect was previously observed near strong current sheets in the models of \cite{Lebreuilly2023b} (see the cartoon illustration in their figure 3). We note that, if the variations of dust-to-gas ratio are small in our case, there is room for an instability enhancing them either by slowing down the collapse (more rotation, more turbulence, stronger magnetic fields) or accelerating coagulation. The latter might be possible when considering other types of turbulence as the one of the Ormel kernel, as shown in \cite{Gong2020,Gong2021}. 

\begin{figure}
  \centering
  \includegraphics[width=\columnwidth]{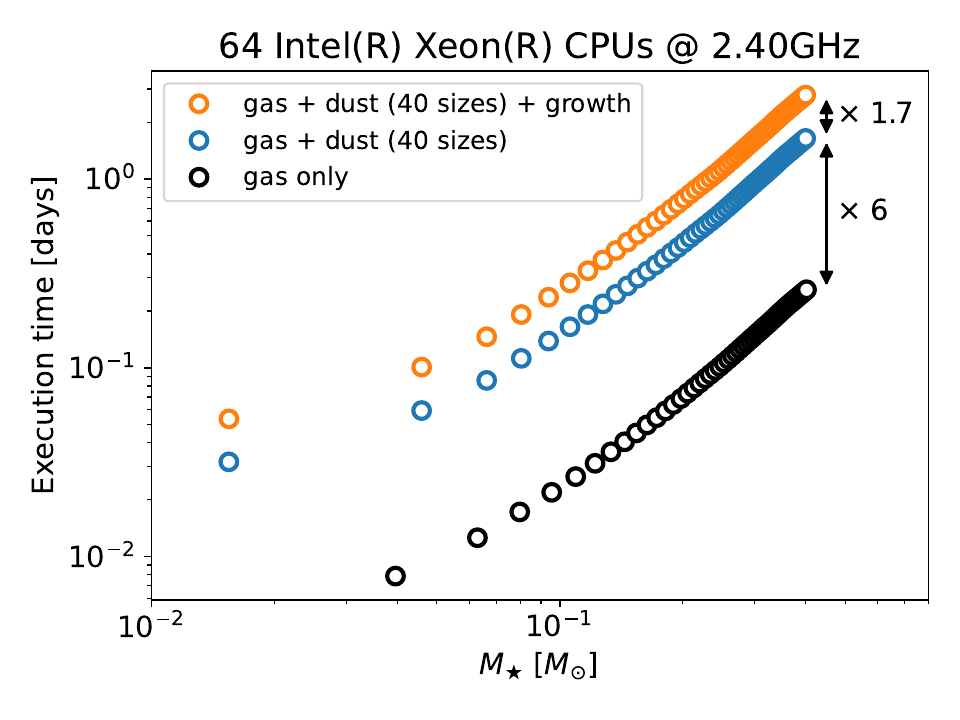}
    \caption{Comparison of the computational time between three simulations for the model $\mathcal{M}_0$: gas only (black), gas with 40 dust sizes (blue), gas with 40 dust sizes and dust growth (orange). Using \texttt{COALA} in \texttt{RAMSES} only increases the global execution time by a factor $1.7$.}
   \label{fig:exec_time}
\end{figure}

Figure~\ref{fig:exec_time} shows the global execution time in function of the mass of the star with the model $\mathcal{M}_0$ for three simulations: gas only, gas with 40 dust sizes, gas with dust growth. By using \texttt{COALA} in \texttt{RAMSES} (orange circles), the computational cost is slightly higher by a factor $1.7$ in the global execution time, compared to the same simulation without dust growth (blue circles). The factor $6$, between the simulation gas only and gas with 40 dust sizes, is coherent with the computational time scaling as $\sqrt{\mathcal{N}}$ as shown in \citet{Lebreuilly2019}. The next goal is to take advantage of the performances of \texttt{COALA} to reduce the number of dust sizes to $\sim 20$ which will reduce the computational cost.

\section{Discussion}
\begin{figure*}
  \centering
  \includegraphics[width=0.8\textwidth,trim={0cm 7cm 0cm 5cm},clip]{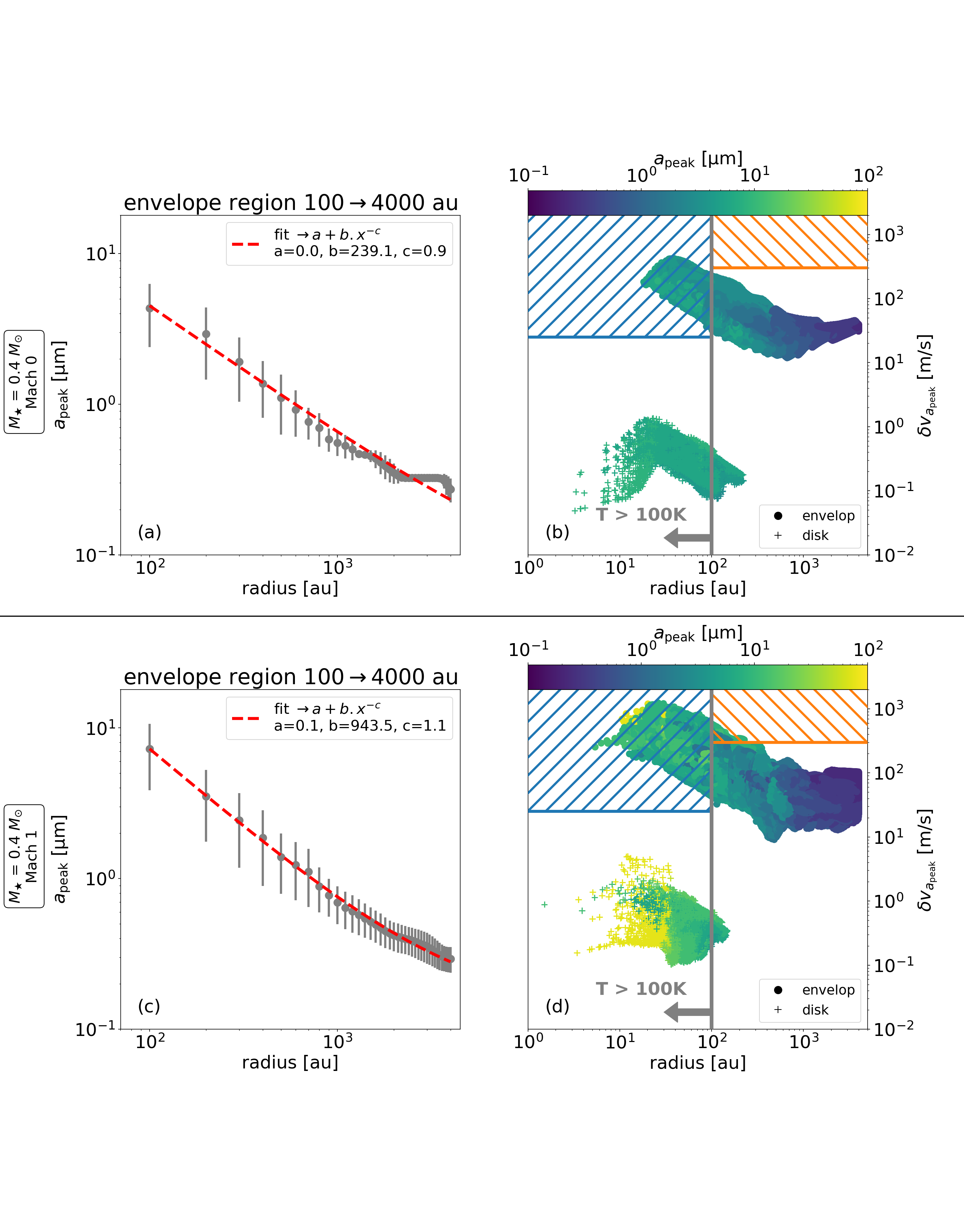}
    \caption{Radial profiles of the dust grain size at the peak of the size distribution \apeak (see text for further details), and the grain-grain differential velocity of grains with size \apeak from turbulence model (see Sect.~\ref{sec:dust_dv_turb}), noted \dvgrain, for the two snapshots shown in Fig.\ref{fig:maps}, models $\mathcal{M}_0$ and $\mathcal{M}_1$ at $t_{0.4 M_{\odot}}$. The radial profiles shown in the left panels are built by computing the mean \apeak in concentric shells from the \apeak map (for further details, see Fig. \ref{fig:shells} and text), they show a clear increase of the mean dust grains size with decreasing envelope radius, up to sizes $\sim 10$\mic. The power-law in model $\mathcal{M}_1$ is higher than for model $\mathcal{M}_0$. This shows that the grain grow faster in model $\mathcal{M}_1$, since the gas turbulence tends to slow down the collapse letting more time for grains to grow. The right panels show the dust grain populations (colored by \apeak) at all envelope radii, and plotted against the \dvgrain they experience at this envelope location (in individual cells in the model). A grey vertical line represents the 100 au radius, approximating the ice-line expected in such solar-type protostars. On each side of this line, orange and blue shaded areas represent the locations where the dust grains are expected to fragment (see text for further details).}
   \label{fig:hist2d}
\end{figure*}

Panels (a) and (c) in Figure~\ref{fig:hist2d} show the radial profiles of the dust grain size at the peak of the distribution, \apeak, in the envelope (100 to 4000 au) for the two models ($\mathcal{M}_0$ and $\mathcal{M}_1$) at $t_{0.4 M_{\odot}}$. The radial profile is built by computing the mean \apeak in concentric shells around the central star (see Appendix~\ref{sec:radial_profil}). The mean value of \apeak is calculated among the cells where the grain-grain differential velocity of grains with size \apeak, noted \dvgrain, is lower than the fragmentation velocity threshold. Both profiles show a clear increase of \apeak, up to $\sim 10\, \mu \rm{m}$, with the decreasing envelope radius. We used the fit function of the form $f(x) \equiv a + bx^{-c}$. The power-law coefficient $c$ goes from $0.9$ for $\mathcal{M}_0$ model to $1.1$ for $\mathcal{M}_1$ model. This power-law is perfectly consistent with the fact that the grain size is $\propto \sqrt{\rho}$ since the density scales roughly as $\propto \frac{1}{r^2}$. We note that the slope is steeper in  $\mathcal{M}_1$, as turbulence (1) slows down the collapse and (2) generates dense filamentary structure far from the center. We observe that both models produce micron-sized grains, while the model $\mathcal{M}_1$ produces slightly larger grains (up to $\sim 8\, \mu$m) in the envelope: we interpret this finding as being due to the turbulence slowing down the collapse and generating dense regions away from the collapse center, which in turns enhance collision rates favoring dust growth.

Figure \ref{fig:hist2d}, panels (b) and (d) show 2D histograms of the grain size \apeak according to the distance to the central star and \dvgrain, the grain-grain differential velocity of grains with size \apeak, from turbulence models \citep[][see Sect.~\ref{sec:dust_dv_turb}]{Ormel2007}. In this model, the level of turbulence is reduced in the disk region (when $n_{\mathrm{H}} > 10^9$) in order to be consistent with the model of $\alpha$-viscosity disk \citep{Birnstiel2010}. Therefore, \dvgrain is reduced by 4 orders of magnitude. The disk region is defined for low values of radii and \dvgrain (cross marker). The point markers, located at higher values of \dvgrain, define the envelope region. 
The realism of predictions on grain sizes, by considering only dust growth, depends on fragmentation limit in order to evaluate if grains can survive. Our modeling does not yet include the fragmentation processes, yet as grains grow in size these are expected to become limiting factors to the dust growth. To mitigate the impact of this lacking physics, we use prescriptions from the literature to evaluate whether the grains formed in our models are expected to fragment. We describe here below the hypotheses made to identify the parameter space where grains from our models are expected to fragment (blue and orange hashed regions in the panels (b) and (d) of Figure \ref{fig:hist2d}).
To compute the fragmentation of the grains in our models, we first assume grains are composed of 0.1 µm monomers. Second, we estimate the fragmentation velocity of the grains by comparing the kinetic energy of the grain and the energy required to break a bond between two monomers as $v_{\mathrm{frag}} \sim 22 \; \mathrm{m\, s^{-1}}$ for silicates and $v_{\mathrm{frag}} \sim 300 \; \mathrm{m\, s^{-1}}$ for ice-coated grains  \citep{Ormel2009,Lebreuilly2023}. Observations of Class 0 disks and hot-corino regions in low-mass protostars suggest a significant fraction of the mass at $r<100$ au may be subject to accretion and viscous heating, with temperatures often larger than 50-100 K \citep{Zamponi2021}. Therefore, while dust grains are expected to be ice-coated in the envelope at radii $r > 100$ au, we assume they are probably bare in the region $r < 100$ au. 
The fragmentation velocities for the two type of grains (ice-coated and bare grains) are reported as orange and blue, respectively, hashed areas in the panels (b) and (d) of Figure \ref{fig:hist2d}. All grains with differential dust velocities \dvgrain (enhanced by gas turbulence from Ormel's model, see Sect.~\ref{sec:dust_dv_turb}) larger than these fragmentation velocities have a high probability to fragment and are thus ignored in the following of our discussion. Note that these differential velocities are computed by taking into account the collision between grains representing the bulk of the dust mass (e.g. with sizes close to \apeak).

We stress that the fragmentation of dust grains in astrophysical environments is not well constrained, as the velocity threshold over which the grain-grain collision will result in grain fragmentation(s) depends on the microphysics of grain such as their size, shape and composition. These grain properties are poorly constrained, and there is no consensus as of which grains survive which collisions. For instance, the analysis of grains collected by the Rosetta mission has shown that 50\% of the aggregates with size a few microns may have fragmented due to the collecting velocities $\simgt 3-7\, \mathrm{m\, s^{-1}}$ \citep{Kim2023}. This velocity threshold is much lower than the theoretical value such as $22\, \mathrm{m\, s^{-1}}$ widely used for bare silicate grains with the same porosity \citep{Ormel2009}. To better constrain the growth of grains up to sizes where their fragmentation is likely in protostellar envelopes, where typical gas velocities are a fraction of $\mathrm{km\, s^{-1}}$, one needs to make hypothesis on the structure of these grown grains, especially their porosity and fractal dimensions. Such a level of modeling is beyond the state-of-the-art: future \texttt{COALA} implementations will consistently treat both dust coagulation and fragmentation using more realistic hypothesis.

To conclude, despite the fact that fragmentation processes are not included yet in the \texttt{COALA} dust evolution model implemented in \texttt{RAMSES}, most of the predicted grains with sizes a few (1-10) microns are expected to survive in the envelope physical conditions. The grains of size ranging from 4.9 to 7.1 µm, which do not enter in the fragmentation regime in the envelope, represent 0.5\% of the total dust mass in the model $\mathcal{M}_1$, and 0.4\% in the model $\mathcal{M}_0$. The precise evolution at disk scales requires a dedicated study with dust growth and fragmentation, and discussions on the hypothesis made on disk model giving low values of \dvgrain seen in Fig.~\ref{fig:hist2d}. Interestingly, if grains grow easily to $\sim 10\,\mu$m in size in the envelopes during the first stages of disk formation, before entering the disks and evolving further, it could explain some properties from the dust in comet 67P \citep{Bentley2016}. Indeed, dust particles collected by the Rosetta probe show that grains larger than $8\,\mu$m are more compact and less elongated than the smaller particles - as would be expected from compaction processes in a disk environment. However, micron-sized grains consist of porous agglomerates that are suggested to be hierarchical on a scale probably covering at least hundreds of nanometres to several micrometres, such as would be expected from the growth processes at work in envelope environments.

\begin{figure*}
    \centering    
    \includegraphics[width=0.9\textwidth,trim={0cm 0cm 0cm 0cm},clip]{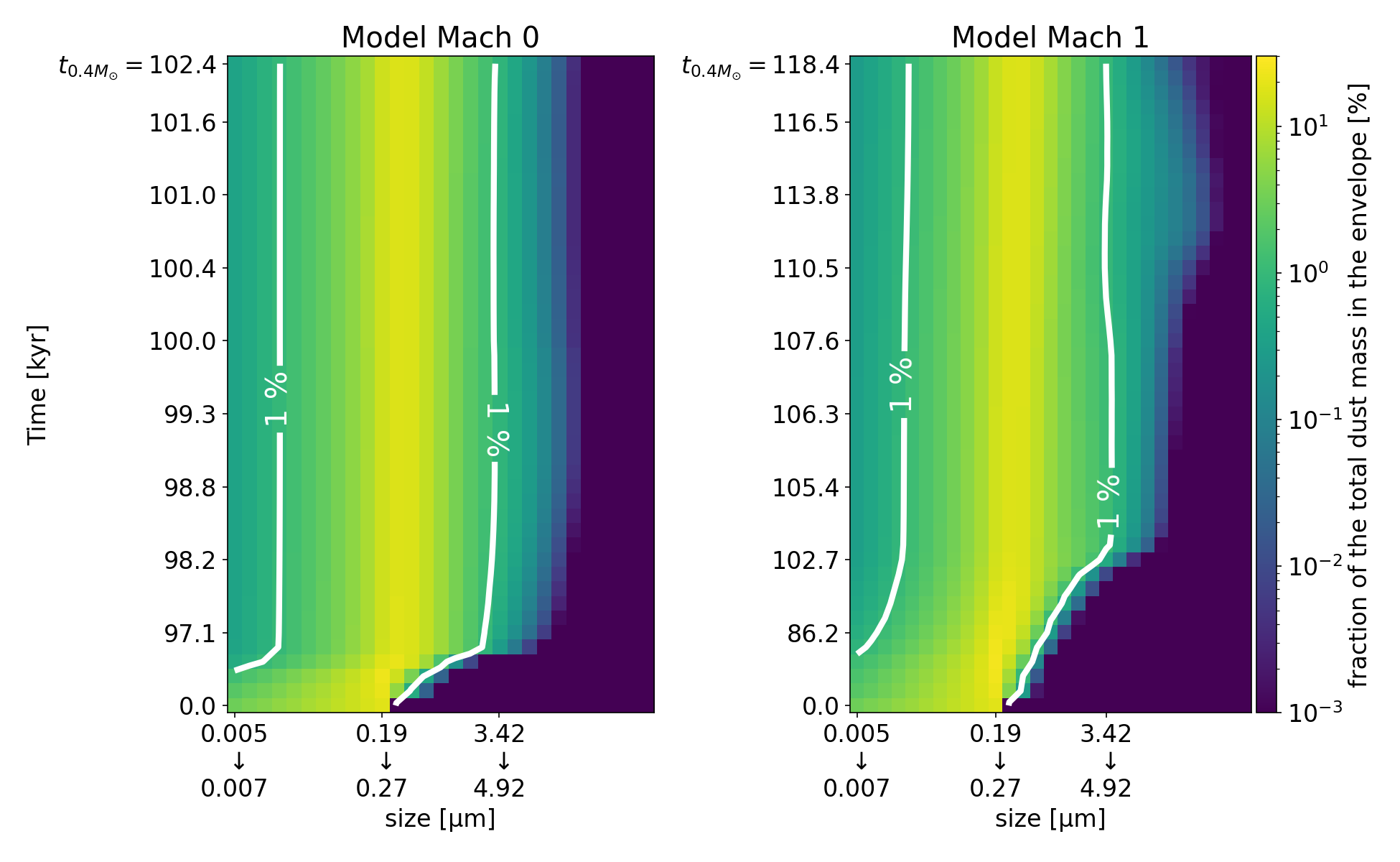}
    \caption{Time evolution of the fraction of the total dust mass in the envelope represented by each grain size bin. Initially the grains from 0.19 to 0.27 µm represent $\sim 20\%$ of the total dust mass. After 100 kyrs, these grains account for $\sim 13\%$ of the total dust mass. A part of the mass is represented by larger grains, such as the grains with size $\sim 3-5$µm which represent steady-state value of $\sim 1\%$ of the total dust mass (white lines).  }
    \label{fig:dust_fraction}
\end{figure*}

Figure~\ref{fig:dust_fraction} shows the evolution of the fraction of the total dust mass represented by each grain size bin in the envelope in the model $\mathcal{M}_1$. Grains can grow easily and very quickly to about 10µm after 100 kyrs in dense envelopes experiencing protostellar collapse. The mass fraction evolution of the grains with size between $\sim 3-5$ µm reaches a steady-state after 100 kyrs with a value $\sim 1\%$ of the total dust mass. 

The evolution of the dust size distribution affects the non ideal MHD resistivities, and thus the magnetic braking and other magnetic effects\citep{Zhao2016,Guillet2020,Silsbee2020,Kawasaki2022,Lebreuilly2023b}. Resistivity evolution are governed by the evolution of the population of small grains. Because our model does not include fragmentation yet, the fraction of very small grains, which drives the AD resistivity, is not accurately described in these current models. In a future study, we will include fragmentation to accurately account for the evolution of MHD resistivities.

Note that, in this study, we did not consider the disruption of grains due to radiative torque from strong radiation fields \citep{Hoang2019,Reissl2024}. This phenomenon may play an important role in the depletion of large grains in the envelope. This fragmentation process will be added in \texttt{COALA} to quantify which grains are rotationaly disrupted to the impact on the dust size distribution in the protostellar envelope, in order to compare to a previous study \citep{LeGouellec2023}.

Our work shows that the dust size distribution rapidly evolves from the initial MRN distribution in protostellar envelopes, due to the coagulation process, as shown in several previous studies \citep{Bate2022,Lebreuilly2023,Vorobyov2025}. 
Therefore, in disks, which material comes from the parent envelope, the initial dust size distribution is most likely not following an MRN, and includes large, micron-sized, dust grains. Such a different dust size distribution may significantly affect the disk evolution, as it will be key into several physical processes such as the disk thermal equilibrium and the chemistry at the grain surfaces, potentially modifying the disk temperature, its scale height and its structure. Moreover, a change in the pristine dust size distribution may also result in different dust size evolution and spatial distribution in the young disks, as well as the coupling of the disk material with the local magnetic field.

\section{Conclusions}
In this work, we performed, for the first time, 3D non-ideal MHD simulations of protostellar collapse that self-consistently accounts for dust dynamics and growth, modeled by the Smoluchowski equation, by coupling the codes \texttt{COALA} and \texttt{RAMSES}. The new code  \texttt{COALA} allows to include dust growth at low numerical cost. This new numerical tool allows to run simulations beyond the current state-of-the-art. Our goal is to predict the grain sizes that expect to survive before fragmentation regime in a collapsing protostellar envelope, during the disk formation stage. Two models with and without initial turbulent Mach are performed up to an evolutionary stage representative of a young embedded protostar. The main findings are detailed in the following:
\begin{itemize}
\item In the envelope environment, dust grains grow rapidly to about $1-10\, \mu \mathrm{m}$ for both models ($\mathcal{M}_0$ and $\mathcal{M}_1$), and survive after 100 kyrs from an initial MRN distribution. 

\item In the disk region, the dust size distribution does not follow an MRN and includes large grains, such as several tens of microns. 

\item Dust enrichment is enhanced by dust growth compared to dust dynamics alone. More models with various kernels are required to see how influential this can be on the dust content of the disk.

\item Dust evolution is highly anisotropic due to the nature of the collapse with preferred locations to produce large grains such as high density structures from a turbulent infall (model $\mathcal{M}_1$) or outflow cavity walls (model $\mathcal{M}_0$). 
\end{itemize}

Dust fragmentation processes from grain-grain collisions and radiative torque are key 
to shape the dust size distribution by counterbalancing the dust growth. Therefore, it is essential to also self-consistently accounts for the dust fragmentation in the 3D MHD simulation in order to predict precisely the dust size distribution locally in the envelope and in the disk. An accurate prediction of the dust size distribution is key for several physical and chemical processes in the protostellar collapse. New 3D simulations of protostellar collapse with dust growth and fragmentation will be performed with the new tool \{\texttt{COALA} + \texttt{RAMSES}\} in a future study.

\begin{acknowledgements}
We thank the referee for providing very useful comments that helped us to improve our manuscript. This work was made possible thanks to the support from the European Research Council (ERC) under the European Union's Horizon 2020 research and innovation program (Grant agreement No. 101098309 - PEBBLES). The simulations were carried out on the Alfven supercomputing cluster of the Commissariat à l’Énergie Atomique et aux énergies alternatives (CEA). Post-processing and data visualization was done using the open source Osyris package.
\end{acknowledgements}
\bibliographystyle{aa}
\bibliography{ref}

%
%

\appendix

\section{Grain-grain differential velocities}
\label{sec:dust_dv}

In this study, the sources of the grain-grain differential velocities are the turbulence, the Brownian motion and the hydrodynamical drift.

\subsection{Differential velocity from gas turbulence}
\label{sec:dust_dv_turb}
The gas turbulence is a source of relative velocity between grains through the drag force. The strength of this turbulent motion depends on the grain size and the Reynolds number $\mathrm{Re}$. \citet{Ormel2007} formulated the expression of the relative velocity between two grains $i$ (the larger grain) and $j$ (the smaller grain) in three regimes which depends on the Stokes number $\St_i \equiv t_{\rm{s},i}/t_{\rm{dyn}}$ where $t_{\rm{dyn}}$ is the dynamical timescale of the system. As shown in \cite{Marchand2021}, we can consider only the intermediate regime in the context of the collapse. In that case, the differential velocity writes 
\begin{equation}
    \Delta v_{\mathrm{turb},i,j} \equiv   \sqrt{\alpha \beta_{i,j}} c_{\mathrm{s}}  \sqrt{\St_i},
\end{equation}
The term $\beta_{i,j}$ is written as
\begin{equation}
    \beta_{i,j} = 3.2 - \left( 1 + \xi_{i,j} \right) + \frac{2}{1+\xi_{i,j}} \left( \frac{1}{2.6} + \frac{\xi_{i,j}^3}{1.6 + \xi_{i,j}} \right),
\end{equation}
where $\xi_{i,j} \equiv \St_i/\St_j$. In this study, the $\alpha$ coefficient, describing the level of turbulence, has a value of $1.5$ in the envelope and $10^{-4}$ in the disk region, defined with a criteria on the number density of hydrogen $n_{\mathrm{H}} > 10^9\; \mathrm{cm^{-3}}$. As in \cite{Tsukamoto2021}, we compute the dynamical time as 
\begin{equation}
t_{\rm{dyn}} = \frac{c_{\mathrm{s}}}{|\vec{a_{\rm{g}}}|},
\end{equation}
$\vec{a_{\rm{g}}}$ being the gravitational acceleration.

\subsection{Brownian motion}
\label{sec:dust_dv_Br}
The differential velocity between two grains of masses $m_i$ and $m_j$ is formulated as
\begin{equation}
    \Delta v_{\mathrm{Br},i,j} \equiv \sqrt{\frac{8 k_{\mathrm{B}}T}{\pi}} \sqrt{\frac{1}{m_i} + \frac{1}{m_j}},
\end{equation}
where $k_{\mathrm{b}}$ is the Boltzmann constant and $T$ the gas temperature. The differential velocity increases as the masses of the grains decrease. Therefore, the dust growth by Brownian motion is efficient for small grains.

\subsection{Hydrodynamical drift}
\label{sec:dust_dv_hydro}
The dynamics of the grains depends on their size due to the drag force from the gas. Therefore, two grains of different sizes can collide. The relative velocity between grains sizes $i$ and $j$ is written as
\begin{equation}
    \Delta v_{\mathrm{drift},i,j} \equiv |\vec{v}_i - \vec{v}_j|.
\end{equation}
This velocity is named hydrodynamical drift velocity.

\section{Benchmarking \texttt{COALA}}
\label{sec:benchmark_coala}

\begin{figure}
    \centering
    \includegraphics[width=\columnwidth,trim={1cm 0cm 3cm 0cm},clip]{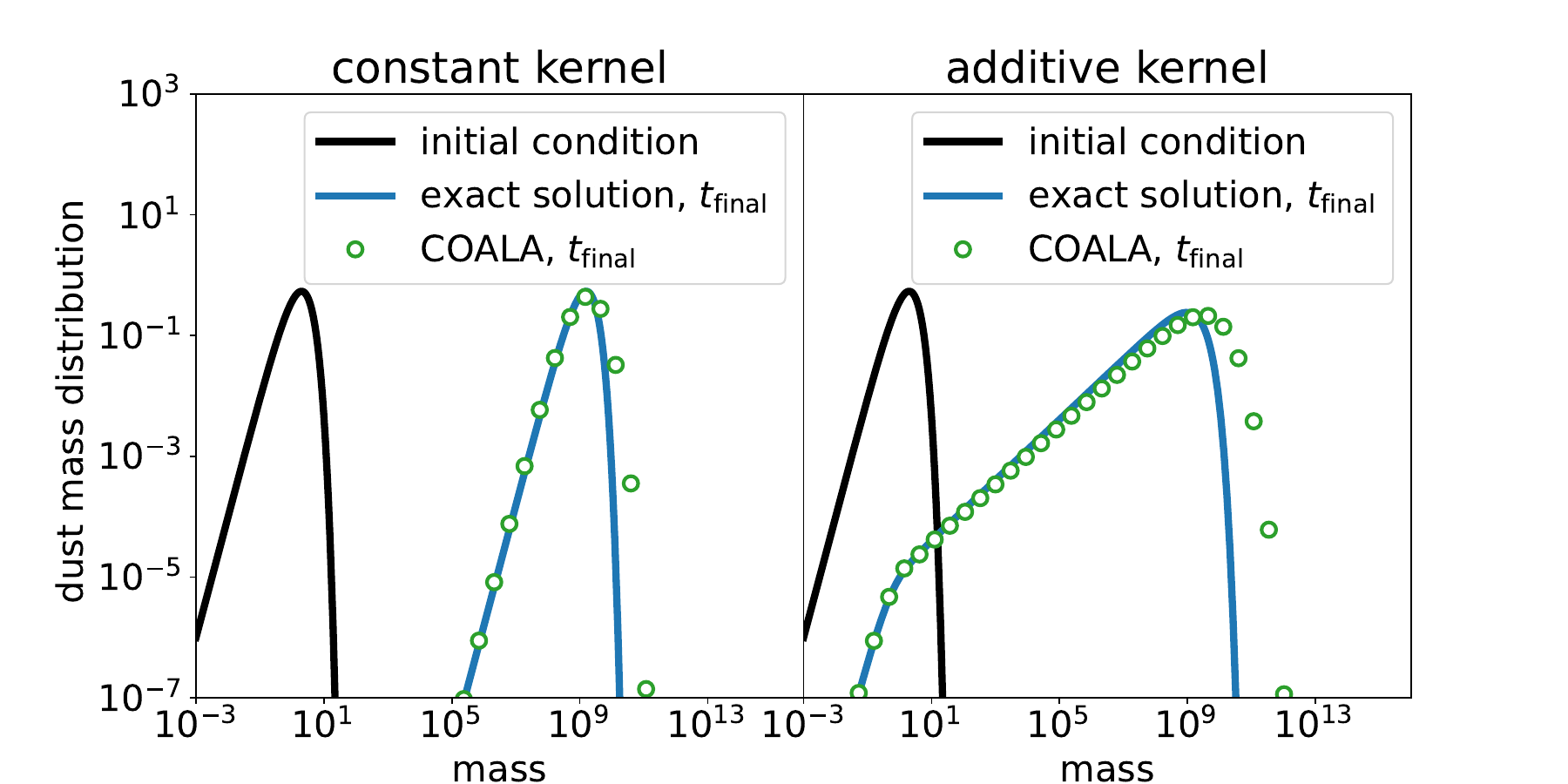}
    \caption{Benchmarks of the code \texttt{COALA} to solve the Smoluchowski coagulation equation (Eq.~\ref{eq:smolu_cons}) with the constant and the additive kernels. In the two plots, the y-axis are the dust grain mass distribution and the x-axis the grain masses in dimensionless unit. The mass range is chosen in order to have 19 orders of magnitude in mass similar to the mass range obtained from $s_{\mathrm{min}}=5 \mathrm{nm}$ to $s_{\mathrm{max}} = 1 \mathrm{cm}$. The black lines are the initial mass distribution. The blue lines are the exact solutions at a final time $t_{\mathrm{final}}$. The green circle are the numerical solutions for which the mass range is sampled in 40 mass bins. The numerical solutions follow accurately the exact solutions.} 
    \label{fig:benchmarks_coala}
\end{figure}

The code \texttt{COALA} is benchmarked with solutions to the Smoluchowski equation, for the constant and additive kernel, in order to check the good accuracy in the setup with the coupling to the code \texttt{RAMSES}. The mass range spans 19 orders of magnitude, which is similar to the one obtained from the size range $s_{\mathrm{min}} = 5\,\mathrm{nm}$ to $s_{\mathrm{max}} = 1\,\mathrm{cm}$. The mass range is discretized in 40 bins. The results are shown in Figure~\ref{fig:benchmarks_coala} where the numerical solutions are in green and the exact ones are in blue. In both tests, the analytical solution is well reproduced up to a final time where grains grown by 3 orders of magnitude in size. We note a slight numerical diffusion at the high mass tail for the additive kernel. This numerical problem is well known from most of the solvers for the Smoluchowski equation. The new code \texttt{COALA} is well designed to drastically reduce this numerical diffusion by taking advantage of high-order approximation \citep{Lombart2021}. In this study, only the piecewise constant approximation is used to keep low numerical costs for the 3D simulations.

\section{Selected region in the simulation for the analysis}
\label{sec:full_map}
\begin{figure*}
    \centering
    \includegraphics[width=0.8\textwidth,trim={0cm 3cm 0cm 3cm},clip]{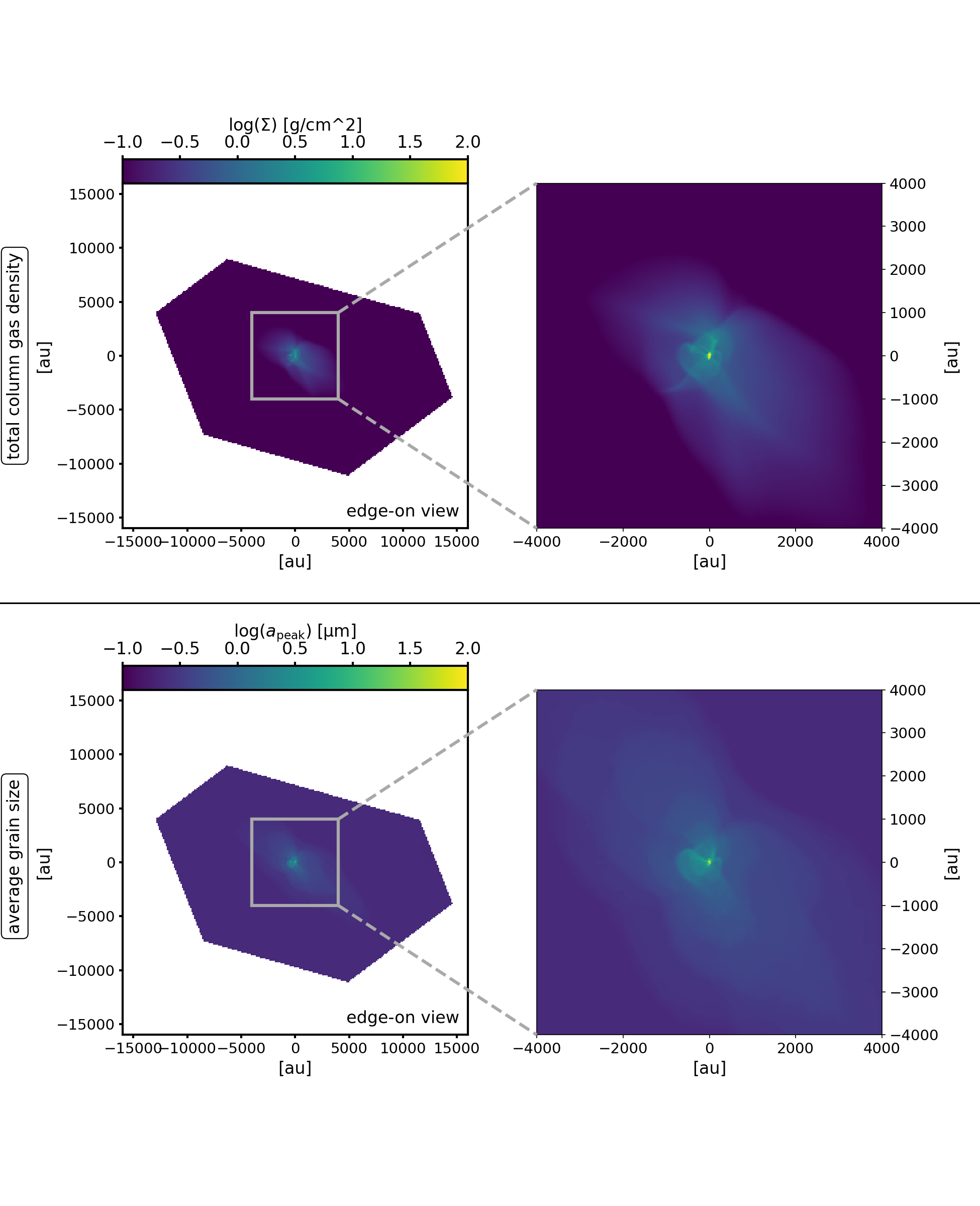}
    \caption{ 
    This shows the selected cube of 8000 au side (grey square) in the simulations of around 32000 au side to perform the analysis in Fig.~\ref{fig:hist2d}. The upper row shows the total column gas density maps and the lower row shows the grain size at the peak of the dust size distribution.  }
    \label{fig:full_size_map}
\end{figure*}

The \texttt{RAMSES} simulations is around 32000 au side. As indicated in \ref{sec:results}, we select 8000 au side cube in order to exclude the impact of the grid boundaries. Figure~\ref{fig:full_size_map} shows the full map and the selected region in the edge-on view. 

\section {Line-of-sight averaged quantities}
\label{sec:av_quant}

The two average quantities shown in the maps and discussed throughout the paper are calculated as:
\begin{equation}
    a_{\mathrm{peak}}\equiv a_{\mathrm{peak},i,j} = \frac{\sum_{k} a_{\mathrm{peak},i,j,k} \;\rho_{i,j,k}}{\sum_{k} \rho_{i,j,k}}, \;\frac{\epsilon}{\epsilon_0} \equiv \frac{\epsilon_{i,j}}{\epsilon_0} = \frac{\sum_{k} \frac{\epsilon_{i,j,k}}{\epsilon_0} \;\rho_{i,j,k}}{\sum_{k} \rho_{i,j,k}},
\end{equation}
where $i,j$ and $k$ are the indices for the spatial coordinates of the cells, for which $k$ is the index of the direction along the line of sight. The quantities $a_{\mathrm{peak},i,j,k}$ and $\epsilon_{i,j,k}$ are the grain size at the peak of the size distribution and the dust-to-gas ratio in a given cell.

\section{Scaling of the peak of the dust size distribution}
\label{sec:scaling}

We demonstrate here a simple relation between the typical grain size $s$ and the gas density $\rho$. We assume a simple spherical collapse such as 
\begin{equation}
    \frac{d \rho}{dt} = \frac{\rho}{ t_{\mathrm{ff}}},
\end{equation}
and a monodisperse formulation of the grain size evolution \citep{Stepinski1996,Tsukamoto2021} 
\begin{equation}
    \frac{d s}{dt} = \frac{s}{3 t_{\mathrm{coag}}},
\end{equation}
where the coagulation timescale writes
\begin{equation}
   t_{\mathrm{coag}} = \frac{m_d}{4 \pi s^2 \rho_d \Delta v}, 
   \label{eq:tcoag}
\end{equation}
and the free-fall timescale writes
\begin{equation}
   t_{\mathrm{ff}} = \sqrt{\frac{3 \pi}{32 G \rho} }.
   \label{eq:tff}
\end{equation}
Combining the two equations yields
\begin{equation}
    \frac{d s}{d \rho} = \frac{s}{3 \rho} \frac{ t_{\mathrm{ff}}}{ t_{\mathrm{coag}}}.
    \label{eq:ds_drho}
\end{equation}
Assuming coagulation by turbulence, we can write
\begin{equation}
    \Delta v = A \sqrt{\frac{t_{\mathrm{s}}}{t_{\mathrm{ff}}}}.
    \label{eq:dv}
\end{equation}
Finally, by injecting Eq.~\ref{eq:tstopkw} with $S_{\mathrm{grain}} \leftarrow s$ and Eqs.~\ref{eq:tcoag},\ref{eq:tff},\ref{eq:dv} into Eq.~\ref{eq:ds_drho}, we obtain 
\begin{equation}
 \frac{d s}{\sqrt{s}}= \left(\frac{3 \pi}{32 G} \right)^{1/4}A \frac{\epsilon_{\mathrm{dg}}}{\sqrt{\rho_{\mathrm{grain}}}} \frac{1}{\sqrt{w_{\mathrm{th}}}}\frac{d\rho}{\rho^{3/4}},
\end{equation}
where $\epsilon_{\mathrm{dg}} \equiv \rho_{\mathrm{dust}}/\rho$ is the dust-to-gas ratio. By integration we finally get
\begin{equation}
 \sqrt{s(t)} -\sqrt{s_0} = 2 \left(\frac{3 \pi}{32 G}\right)^{1/4}A \frac{\epsilon_{\mathrm{dg}}}{\sqrt{\rho_{\mathrm{grain}}}} \frac{1}{\sqrt{w_{\mathrm{th}}}} \left[\rho(t)^{1/4}-\rho_0^{1/4} \right],
\end{equation}
or 
\begin{equation}
 s(t) =\left( 2 \left(\frac{3 \pi}{32 G}\right)^{1/4}A \frac{\epsilon_{\mathrm{dg}}}{\sqrt{\rho_{\mathrm{grain}}}} \frac{1}{\sqrt{w_{\mathrm{th}}}} \left[\rho(t)^{1/4}-\rho_0^{1/4} \right] + \sqrt{s_0}\right)^2.
\end{equation}
Assuming a small initial size and density yields
\begin{equation}
 s(t) =4 \left(\frac{3 \pi}{32 G}\right)^{1/2}A^2 \frac{\epsilon_{\mathrm{dg}}^2}{{\rho_{\mathrm{grain}}}} \frac{1}{{w_{\mathrm{th}}}}\sqrt{\rho}.
\end{equation}
We note that $A \simeq \sqrt{2 \alpha} c_s$ and $w_{\mathrm{th}}=\sqrt{\frac{8}{\pi \gamma}} c_s$. We finally get
\begin{equation}
 s(t) \simeq  \sqrt{8 \pi \gamma } \left(\frac{3 \pi}{32 G}\right)^{1/2} \alpha c_s \frac{\epsilon_{\mathrm{dg}}^2}{{\rho_{\mathrm{grain}}}} \sqrt{\rho}
\end{equation}
Or 
\begin{eqnarray}
 s(t)&\simeq &0.004 ~\mathrm{cm}  \left(\frac{\alpha}{1}\right) \left(\frac{T}{10 K}\right)^{1/2} \left(\frac{\epsilon_{\mathrm{dg}}}{0.01}\right)^2 \left(\frac{\rho_{\mathrm{grain}}}{2.3 ~\mathrm{g}~\mathrm{cm}^{-3} }\right)^{-1} \nonumber\\&&\left(\frac{\rho}{10^{-13}~\mathrm{g}~\mathrm{cm}^{-3} }\right)
\end{eqnarray}
We point out that, as shown in \cite{Lebreuilly2020}, the Stokes number St, which is the ratio between the stopping time and the free-fall time scales as $\propto \frac{s}{\sqrt{\rho}}$. This means that this quantity is roughly constant in the isothermal regions of the collapse when dust growth is included.

\section{Model $\mathcal{M}_1$ with and without dust growth}
\label{sec:M1_g_ng}

\begin{figure*}
    \centering
    \includegraphics[width=0.9\textwidth,trim={0cm 3cm 0cm 0cm},clip]{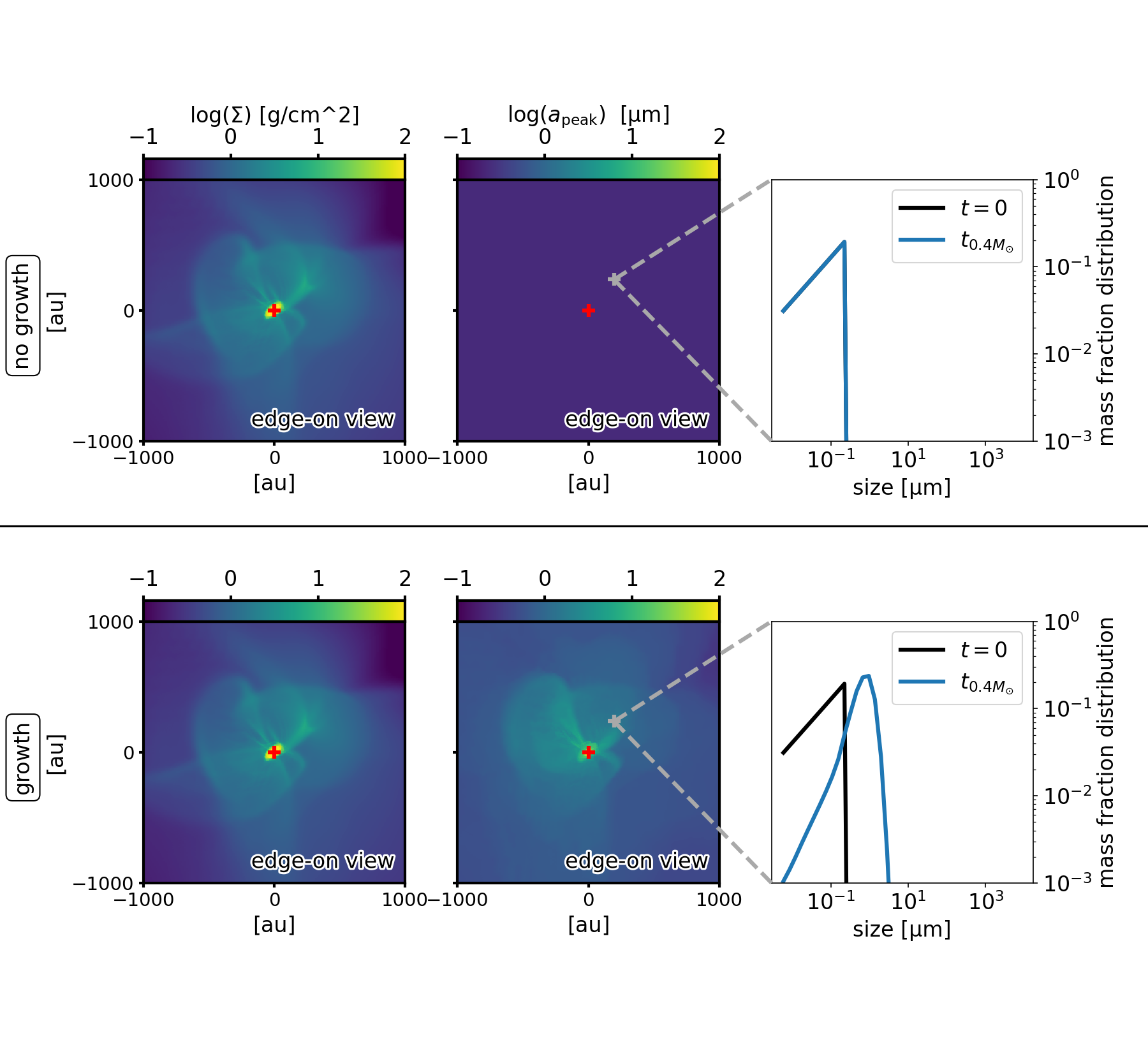}
    \caption{ This shows the comparison without dust growth (upper row) and with dust growth (lower row) for the model $\mathcal{M}_1$ with $M_{\star}=0.4$\msun. The red cross gives the position of the star. The left and middle columns show the total column gas density $\Sigma$ and the average grain size at the peak of the dust size distribution $a_{\mathrm{peak}}$ similarly to Fig.~\ref{fig:maps}. For the model without growth, the $a_{\mathrm{peak}}$ is the maximum size of the MRN distribution ($\sim$ 250\, nm). The right column shows the evolution of the dust size distribution in a cell in the simulation.}
    \label{fig:M1_g_ng}
\end{figure*}

In Figure~\ref{fig:M1_g_ng}, we compared simulations of the model $\mathcal{M}_1$ with (bottom row) and without (top row) dust growth. The total column density $\Sigma$ and \apeak are defined in Sect.~\ref{sec:results}. The dust growth highlights the anisotropy of the dust evolution, where grains grow faster in some specific regions, i.e. structures due to turbulent infall in the model $\mathcal{M}_1$. We note that the initial MRN distribution evolved locally in the simulation due to the dust growth. 

\section{Radial profile of the dust size distribution}
\label{sec:radial_profil}
\begin{figure*}
    \centering
    \includegraphics[width=\textwidth,trim={2cm 5cm 2cm 3cm},clip]{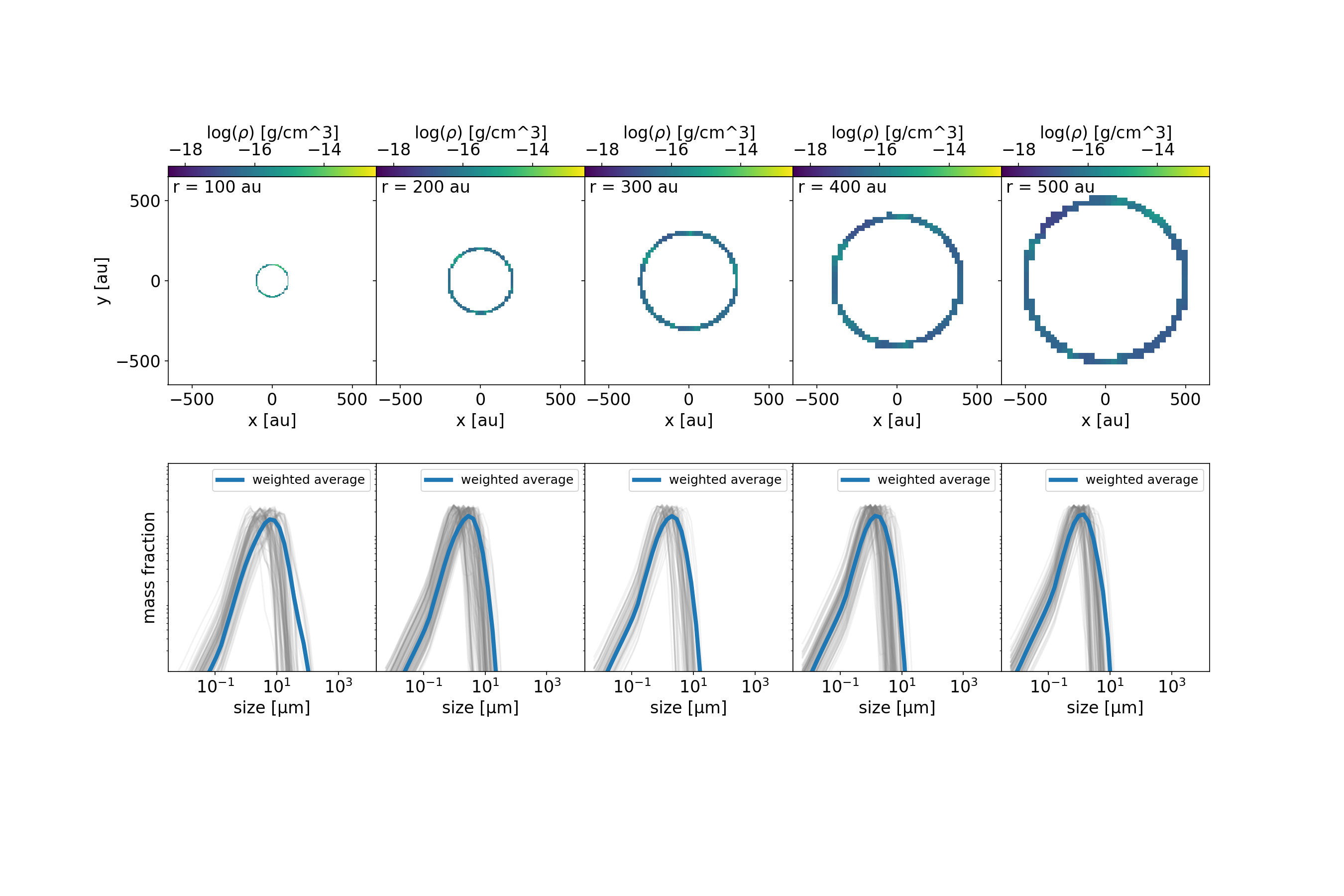}
    \caption{This shows the radial dependency of the dust grain size distribution, from the model $\mathcal{M}_1$ with $M_{\star}=0.4$\msun. Concentric shells of varying radii from 100 to 4000 au (as shown in the upper panels) are used to compute the mean dust grain size distributions shown in the lower panels. The dust grains size distribution is represented by the mass fraction distribution (see text for details). Here only the five first concentric shells with radius [100,200,300,400,500] au are shown in upper panel with the corresponding average dust grain size distribution weighted by the gas density (blue line) in the lower panel. The dust grain size distributions in each cell defining the shell are shown (light grey lines) in the lower panel.}
    \label{fig:shells}
\end{figure*}

The radial profile of the dust size distribution in the 3D simulation is built by computing the mean of the dust size distribution in selected concentric shells around the star. The width of the shells is defined as 10\% of the radius in order to capture enough cells to compute the mean. Figure~\ref{fig:shells} shows the shells with radius [100,200,300,400,500] (top row) with the corresponding average dust size distribution (bottom row). From these average dust size distributions, we compute the mean grain size \apeak at the peak of the distribution.

\end{document}